\documentclass[english,aps,pra,superscriptaddress,twocolumn]{revtex4}
\usepackage[T1]{fontenc}
\usepackage[latin9]{inputenc}
\usepackage{graphicx}
\usepackage{babel,bm}
\usepackage{amsfonts}
\usepackage{amsmath}
\usepackage{amssymb}
\usepackage{graphicx}
\usepackage{color}
\usepackage{epstopdf}
\usepackage{txfonts}

\newcommand{\ket}[1]{\left| #1 \right\rangle}
\newcommand{\bra}[1]{\left\langle #1 \right|}

\definecolor{Blue}{rgb}{0,0,1}
\definecolor{Red}{rgb}{1,0,0}
\definecolor{Green}{rgb}{0,1,0}
\definecolor{Purp}{rgb}{.2,0,.2}
\definecolor{white}{rgb}{1,1,1}

\begin{document}
\title{Experimental reconstruction of work distribution and verification of fluctuation relations at the full quantum level}
\author{Tiago Batalh\~{a}o}
\affiliation{Centro de Ci\^{e}ncias Naturais e Humanas, Universidade Federal do ABC,
R. Santa Ad\'{e}lia 166, 09210-170 Santo Andr\'{e}, S\~{a}o Paulo, Brazil}
\author{Alexandre M. Souza}
\affiliation{Centro Brasileiro de Pesquisas F\'{i}sicas, Rua Dr. Xavier Sigaud 150,
22290-180 Rio de Janeiro, Rio de Janeiro, Brazil}
\author{Laura Mazzola}
\affiliation{Centre for Theoretical Atomic, Molecular and Optical Physics, School
of Mathematics and Physics, Queen\textquoteright{}s University, Belfast
BT7 1NN, United Kingdom}
\author{Ruben Auccaise}
\affiliation{Centro Brasileiro de Pesquisas F\'{i}sicas, Rua Dr. Xavier Sigaud 150,
22290-180 Rio de Janeiro, Rio de Janeiro, Brazil}
\author{Roberto S. Sarthour}
\affiliation{Centro Brasileiro de Pesquisas F\'{i}sicas, Rua Dr. Xavier Sigaud 150,
22290-180 Rio de Janeiro, Rio de Janeiro, Brazil}
\author{Ivan S. Oliveira}
\affiliation{Centro Brasileiro de Pesquisas F\'{i}sicas, Rua Dr. Xavier Sigaud 150,
22290-180 Rio de Janeiro, Rio de Janeiro, Brazil}
\author{John Goold}
\affiliation{The Abdus Salam International Centre for Theoretical Physics, 34014 Trieste, Italy}
\affiliation{Clarendon Laboratory, University of Oxford, Parks Road, Oxford OX1
3PU, United Kingdom}
\affiliation{Department of Physics, University College Cork, Cork, Ireland}
\author{Gabriele De Chiara}
\affiliation{Centre for Theoretical Atomic, Molecular and Optical Physics, School
of Mathematics and Physics, Queen\textquoteright{}s University, Belfast
BT7 1NN, United Kingdom}
\author{Mauro Paternostro}
\affiliation{Centre for Theoretical Atomic, Molecular and Optical Physics, School
of Mathematics and Physics, Queen\textquoteright{}s University, Belfast
BT7 1NN, United Kingdom}
\affiliation{Institut f\"ur Theoretische Physik, Albert-Einstein-Allee 11, Universit\"at Ulm, D-89069 Ulm, Germany}
\author{Roberto M. Serra}
\affiliation{Centro de Ci\^{e}ncias Naturais e Humanas, Universidade Federal do ABC,
R. Santa Ad\'{e}lia 166, 09210-170 Santo Andr\'{e}, S\~{a}o Paulo, Brazil}

\begin{abstract}
 Research on the out-of-equilibrium dynamics of quantum systems has so far produced important statements on the thermodynamics of small systems undergoing quantum mechanical evolutions
 . Key examples are provided by the Crooks and Jarzynski relations
 : taking into account fluctuations in non-equilibrium dynamics, such relations connect equilibrium properties of thermodynamical relevance with explicit non-equilibrium features. Although the experimental verification of such fundamental relations in the classical domain has encountered some success
 , their quantum mechanical version requires the assessment of the statistics of work performed by or onto an evolving quantum system, a step that has so far encountered considerable difficulties in its implementation due to the practical difficulty to perform reliable projective measurements of instantaneous energy states
 . In this paper, by exploiting a radical change in the characterization of the work distribution at the quantum level
 , we report the first experimental verification of the quantum Jarzynski identity and the Tasaki-Crooks relation
 following a quantum process implemented in a Nuclear Magnetic Resonance (NMR) system. Our experimental approach has enabled the full characterisation of the out-of-equilibrium dynamics of a quantum spin in a statistically significant way, thus embodying a key step towards the grounding of quantum-systems thermodynamics.
\end{abstract}
\maketitle

The verification and use of quantum fluctuation relations~\cite{Tasaki,Crooks,Jarzynski} requires the design of experimentally feasible strategies for the determination of the work distribution following a process undergone by a system. In the quantum regime, the concept of work done by or on a system needs to be reformulated~\cite{Talkner} so as to include {\it ab initio} both the inherent non-deterministic nature of quantum dynamics and the effects of quantum fluctuations. In this sense, work acquires a meaning only as a statistical expectation value $\langle W\rangle=\int\,W\,P(W)\,dW$ that accounts for the possible trajectories followed by a quantum system across its evolution, as formalised by the associated work probability distribution
$P(W)=\sum_{n,m} p^0_n\;  p^\tau_{m \vert n} \delta\left[W-(\overline\epsilon_m-\epsilon_n)\right]$. In order to understand this expression, let us consider a quantum system initially at equilibrium at temperature $T$ and undergoing a quantum process that changes its Hamiltonian as $\hat{\cal H}(0)\to\hat{\cal H}(\tau)$ within a time period $\tau$. Then, $p^0_{n}$ is the  probability to find the system in the eigenstate $\ket{n(0)}$ of $\hat{\cal H}(0)$ (with energy $\epsilon_n$) at the start of the protocol, while $p^\tau_{m\mid n}=|\langle m(\tau)\vert\hat{\cal U}\vert n(0)\rangle|^2$ is the conditional probability to find it in the eigenstate $\ket{m(\tau)}$ of $\hat{\cal H}(\tau)$ (with energy $\overline\epsilon_m$) if it was in $\ket{n(0)}$ at $t=0$ and evolved under the action of the propagator $\hat{\mathcal{U}}$. $P(W)$ encompasses the statistics of the initial state (given by $p_n^0$) and the fluctuations arising from quantum measurement statistics (given by $p^\tau_{m \vert n}$). One can define a {\it backward} process that, starting from the equilibrium state of the system associated with $\hat{\cal H}(\tau)$ and temperature $T$, implements the protocol $\hat{\cal H}(\tau)\to\hat{\cal H}(0)$ and thus inverting the control sequence of the process itself. 
The work probability distributions for the forward ($F$) and backward ($B$) processes allow for the formulation of quantum fluctuation theorems of key foundational importance~\cite{campisi,Esposito}.
 
 It is often convenient to work with the Fourier transform of the work distribution, or work characteristic function. For the $F$ process, this is defined as $\chi(u)=\int{P_{F}(W)e^{-iuW}}dW$ and takes the form
 \begin{equation}
\label{eq:workcharfunc}
\begin{aligned}
\chi_F(u) & =\sum_{m,n}p^0_{n}p^\tau_{m\mid n}e^{iu(\overline{\epsilon}_{m}-\epsilon_{n})}
 =\operatorname{Tr}[ (\hat{\cal U}e^{-iu\,\hat{\cal H}(0)}) \rho_0 ( e^{-iu\,\hat{\cal H}(\tau)}\hat{\cal U})^{\dagger}]
\end{aligned}
\end{equation}
with $\rho_0$ the initial equilibrium state of the system (at a given temperature). The characteristic function of the $B$ process is defined analogously. Needless to say, $P_{\alpha}(W)$ and $\chi_{\alpha}(u)$ ($\alpha=F,B$) bring about the same statistical information and they are equally useful. 

The experimental verification of such fundamental relations in the classical domain has found some success~\cite{Collin,Liphardt,Saira,Toyabe,Douarche}. Albeit a few proposals for the measurement of the quantum statistics of work have been made~\cite{Huber,Heyl}, including an ingenious calorimetric one~\cite{Pekola}, the experimental reconstruction of either $P_\alpha(W)$ or $\chi_\alpha(u)$ has been hindered by the technical difficulties inherent in the determination of the conditional probabilities $p^\tau_{m|n}$.  Recently, however, an alternative approach to this problem has been devised, based on well-known interferometric schemes of the estimation of phases in quantum systems, which bypasses the necessity of direct projective measurements on the instantaneous state for the system and paves the way to the faithful reconstruction of the characteristic function of a process~\cite{Dorner,Mazzola} (see Ref.~\cite{CampisiNew} for an interesting development of the original proposal). Here, we exploit such protocol to reconstruct the statistics of work on a spin-$1/2$ system undergoing a quantum non-adiabatic evolution, and thus achieve sufficient information on its out-of-equilibrium features to faithfully verify both the Tasaki-Crooks and Jarzynksi identities with high statistical significance. To the best of our knowledge, this is the first experimental assessment of fluctuation relations fully in the quantum regime.

\begin{figure}
\includegraphics[width=0.88\columnwidth]{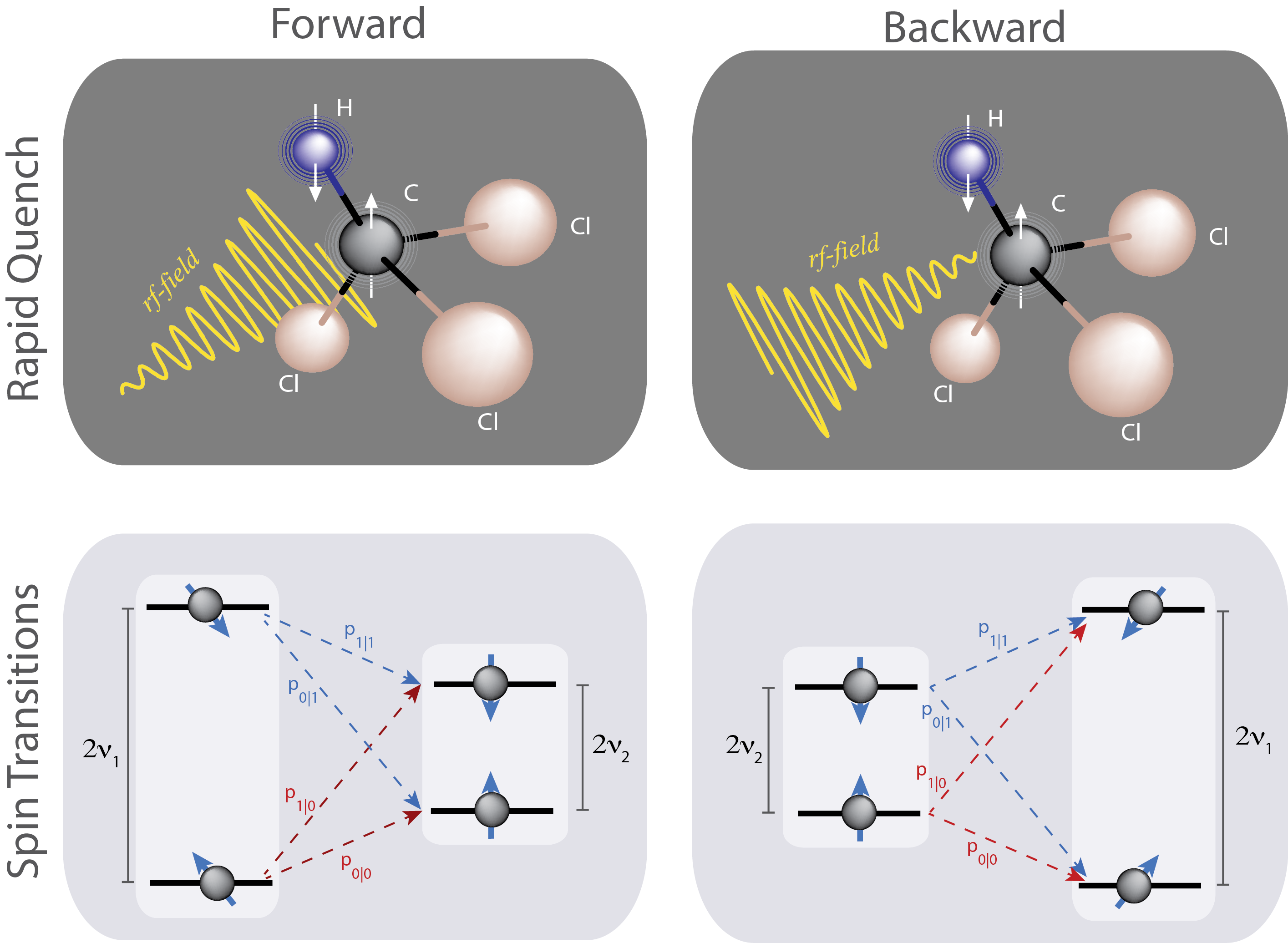}
\caption{{Forward and backward processes}. Upper panels: Quench of the rf-field on the $^{13}$C nuclear spin of a chloroform molecule.  Lower panels: Sketch of the energy spectrum and possible transitions during the quenched dynamics. Each left (right) panel is relative to the forward (backward) process.}
\label{fig:process}
\end{figure}   

Our experiment was carried out using  liquid-state NMR-spectroscopy of the $^{1}$H and $^{13}$C nuclear spins of a chloroform-molecule sample (cf. Fig.~\ref{fig:process}). This system can be regarded as an ensemble of identical, non-interacting, spin-$1/2$ pairs. The rest of the molecule, indeed, can be disregarded, providing mild environmental effects  that, within the time-span of our experiments, are inessential to our results. The main sources of imperfections are identified later on in this paper. 

\begin{figure}
\includegraphics[width=0.97\columnwidth]{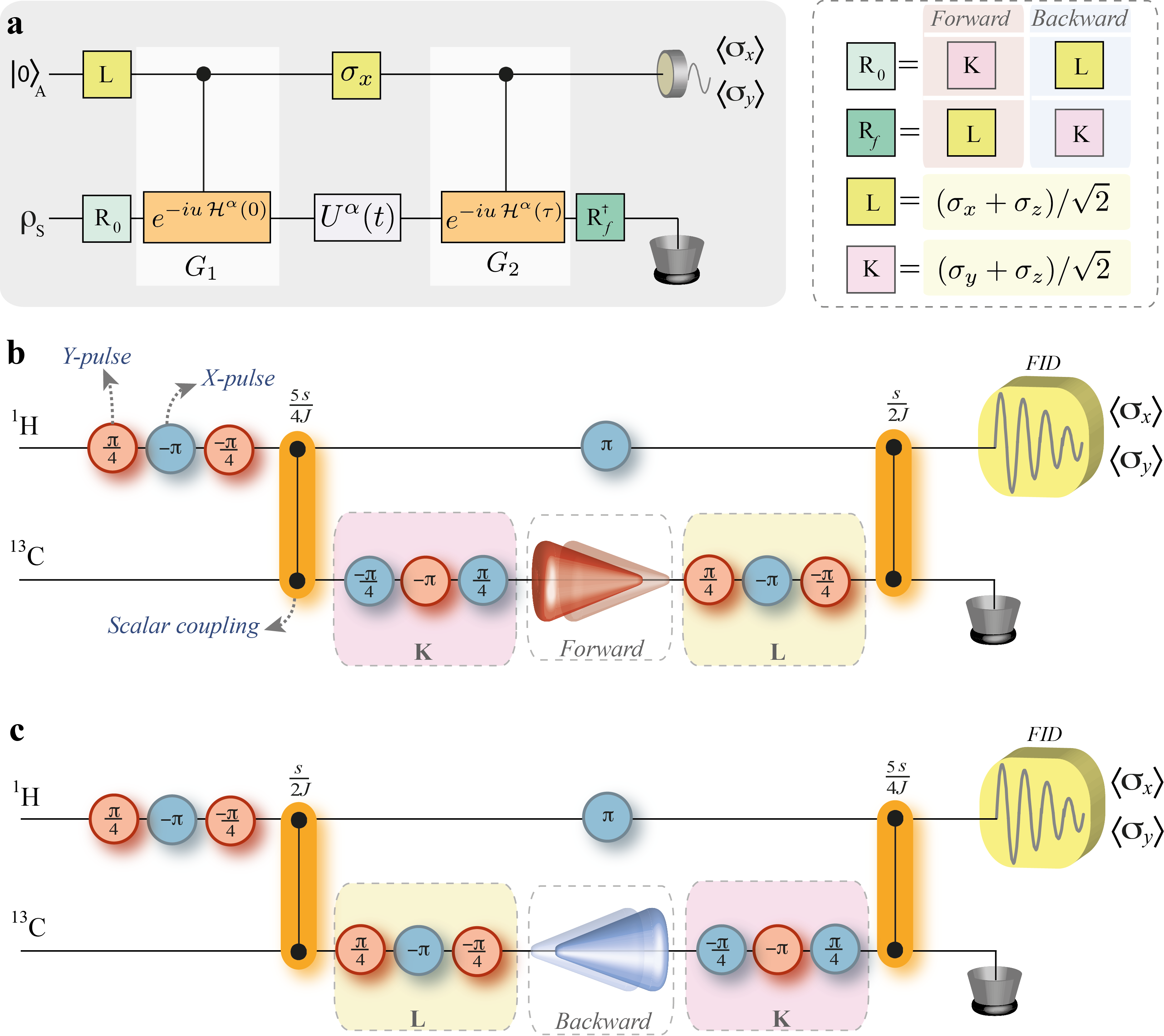}
\caption{{NMR pulse-sequence for the reconstruction of the work distribution}. Panel \textbf{a}: General quantum algorithm for the interferometric reconstruction of $\chi_\alpha(u)$ ($\alpha=F,B$) inspired by the proposals in Refs.~\cite{Dorner,Mazzola}. We show both the conditional joint gates given in Eqs.~\eqref{eq:conditionals} and the single-spin operations needed to complete the protocol. Here, $\rho_S$ is a generic initial state of a driven system (the $^{13}$C nuclear spin in our experiment), while $\ket{0}_A$ is an initial preparation for the ancilla (the $^{1}$H nuclear spin). Panel \textbf{b} (Panel \textbf{c}): The circuit for the $F$ ($B$) process. The blue (red) circles represent transverse rf-pulses in the $x$ ($y$) direction that produce rotations by the displayed angle. Evolutions under the interaction $\hat{\cal H}_J=2\pi J \hat\sigma^H_z \hat\sigma^C_z$ (with $J \approx 215.1$~Hz and for a set amount of time) are represented by two-qubit gates (in orange). The time-length of the coupling is set by the angle $s$, which is related to conjugate variable $u$ in Eq. (\ref{eq:workcharfunc}) by $s = 2\pi \nu_1 u $. The pulse sequence identified by $L$ [$K$] produces the Hadamard gate $(\hat\sigma_x^{C} + \hat\sigma_z^{C} )/\sqrt{2}$ [$(\hat\sigma_y^{C}+\hat\sigma_z^{C} )/\sqrt{2}$].}
 \label{fig:circuit}
\end{figure}

The $^{13}$C nuclear spin plays the role of a driven system, while the $^{1}$H one embodies an ancilla that will be instrumental to the reconstruction of $\chi_\alpha(u)$. The process implemented in our experiment consists of a rapid change in a time-modulated radio frequency (rf) field at the frequency of the $^{13}$C nuclear spin. Formally, this can be described by the following time-dependent Hamiltonian (in the rotating frame and for the $F$ process only)
\begin{align}
 \hat{\cal H}^{F}(t) &= 2\pi\hbar \nu\left( t \right) 
 \left( \hat\sigma_x^{C}\sin\frac{\pi t}{2\tau}  + \hat\sigma_y^{C}\cos\frac{\pi t}{2\tau}  \right),
 \label{eq:quench}
\end{align} 
where $\hat{\sigma}^C_{x,y,z}$ are the Pauli operators for the $^{13}$C spin and $\nu(t)=\nu_1 \left(1-t/{\tau}\right) + \nu_2 t/{\tau}$ is a linear ramp (taking an overall time $\tau= 0.1${ ms}) of the rf field frequency, from $\nu_1 = 2.5\text{ kHz}$ to $\nu_2 = 1.0\text{ kHz}$, $t \in [0,\tau]$. The chosen value of $\tau$ is smaller than the evolution time of the Hamiltonian in the non-adiabatic regime. The reverse quench (realising the $B$ process) is described by $\hat{\cal H}^{B}(t)=-\hat{\cal H}^{F}(\tau-t)$. 
Both processes are sketched in Fig. \ref{fig:process}. In order to reconstruct the work distribution of both the $F$ and $B$ process, we make use of the proposals put forward in Refs.~\cite{Mazzola, Dorner}, which rely on the Ramsey-like interferometric scheme illustrated in Fig.~\ref{fig:circuit}\textbf{a}. 
Through a series of one- and two-body operations, this protocol maps the characteristic function of the work distribution for a system $S$ (the $^{13}$C nuclear spin in our case) prepared in the (equilibrium) state $\rho_S$ and undergoing the process $\hat{\cal H}^{\alpha}(0)\to\hat{\cal H}^{\alpha}(\tau)$ onto the transverse magnetisation of an ancillary system $A$ (the $^{1}$H nuclear spin), initialised in $\ket{0}_A$.  
     
In the first step of our experiment, we used spatial averaging methods to prepare the $^{1}$H-$^{13}$C nuclear-spin pair in the state equivalent to $\rho_{HC}^0=\ket{0}\bra{0}_H \otimes\rho_C^0$, with $\rho_C^0=e^{-\beta \hat{\cal H}^\alpha(0)}/Z_0$ an equilibrium state of the $^{13}$C nuclear spin at temperature $T$. We have introduced the logical states of the $^{1}$H nuclear spin $\{\ket{0},\ket{1}\}_H$, the inverse temperature $\beta= (k_B T)^{-1}$ ($k_B$ is the Boltzmann constant), and the partition function $Z_0=\mathrm{Tr}[e^{-\beta \hat{\cal H}^\alpha(0)}]$. 

\begin{figure*}
\includegraphics[width=0.80\textwidth]{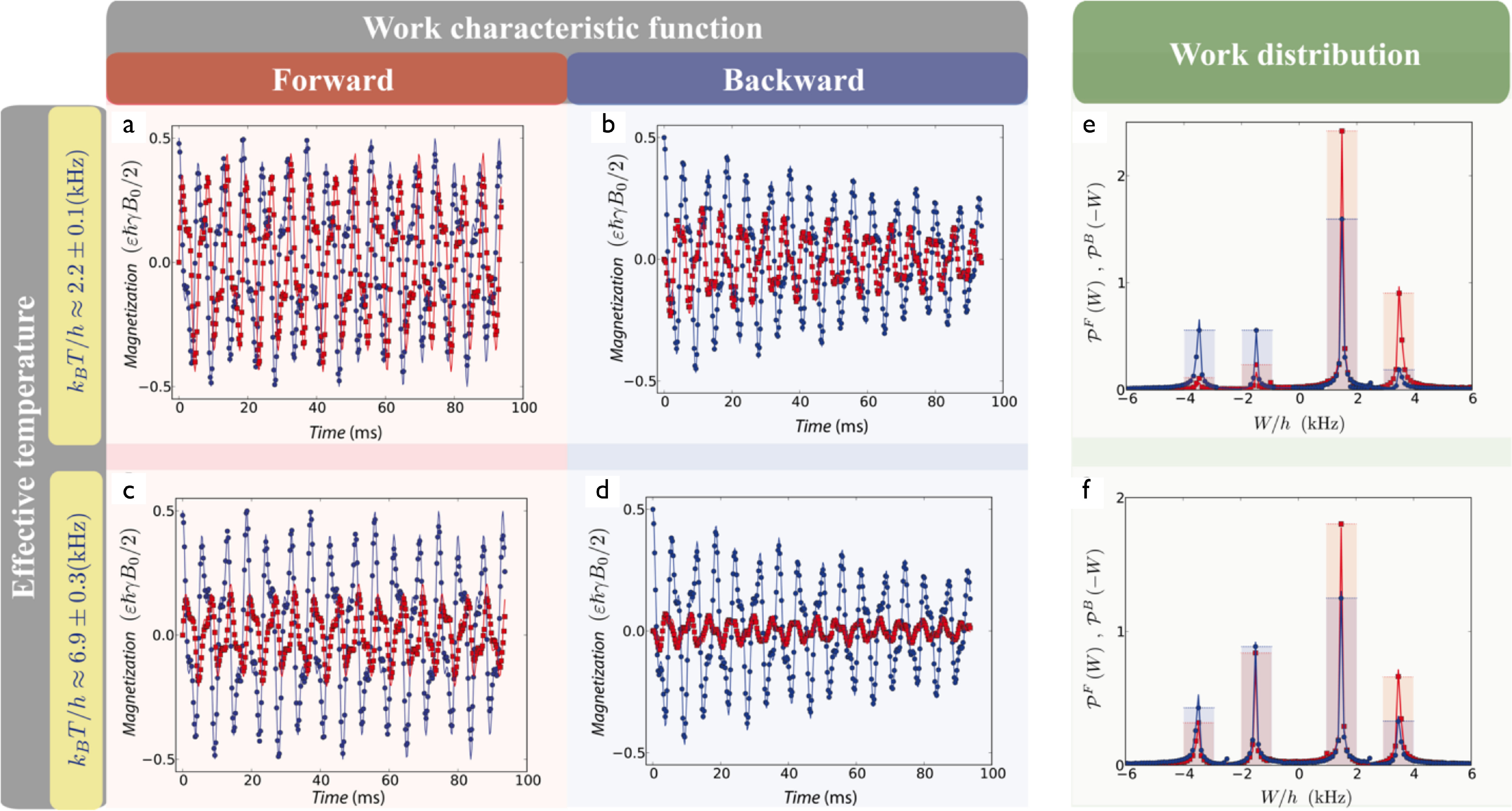}
\caption{{Experimental characteristic function and work distribution for the forward and backward processes}. Panels {\bf a}-{\bf d}: The blue circles (red squares) show the normalized experimental data for the $x$-component ($y$-component) of the $^1$H transverse magnetisation at two different values of the temperature of the system. We study both the forward and the backward process. The solid lines show Fourier fittings, which are in excellent  agreement with the theoretical simulation of the process. The horizontal axis is the evolution time for the shortest coupling, $s/(2J)$ in the pulse-sequence implemented for the reconstruction of $\chi_F(u)$. The error bars in the magnetisation measurement are smaller than the size of the symbols, and are not shown. The uncertainty in the temperature is due to finite precision in the initial state preparation. See Appendix for the definition of $\epsilon,\gamma$ and $B_0$. Panels {\bf e} and {\bf f}: We show the modulus of the inverse Fourier transform of the transverse magnetisation plotted in panels {\bf a}-{\bf d}. The experimental points for the distribution corresponding to the forward (backward) process are presented as red squares (blue circles). The horizontal axis was inverted for the backward process. The experimental data are well fitted by a sum of four Lorentzian peaks centred at $\pm1.5\pm0.1$~kHz and $\pm3.5 \pm 0.1$~kHz  (solid lines), in agreement with the theoretical expectation [for both $\hat{\cal H}^{F(B)}(0)$ and $\hat{\cal H}^{F(B)}(\tau)$] that predict the peaks location to be at $\pm (\nu_1\pm\nu_2)$. 
The amplitudes of the peaks, from the leftmost to the rightmost  in each panel, are proportional to the probabilities $p^{0}_1 p^\tau_{0|1}$, $p^{0}_1 p^\tau_{1|1}$,  $p^{0}_0 p^\tau_{0|0}$, $p^{0}_0 p^\tau_{1|0}$ respectively.}
\label{fig:mag}
\end{figure*}

The structure of Eq. (\ref{eq:workcharfunc}) suggests that one can reconstruct the characteristic function using simple single-spin operations and only two joint gates, each controlled by the ancilla state, and reading
\begin{equation}
\label{eq:conditionals} 
\begin{aligned}
{\hat G}_1 &\equiv \ket{0}\bra{0}_H \otimes e^{-i u \,\hat{\cal H}^\alpha\left(0\right)} + \ket{1}\bra{1}_H \otimes \hat\openone^{C}, \\
{\hat G}_2 &\equiv \ket{0}\bra{0}_H \otimes \hat\openone^{C} + \ket{1}\bra{1}_H \otimes e^{-i u \hat{\cal H}^\alpha\left(\tau\right)}.
\end{aligned}
\end{equation}
The full sequence of operations needed to reconstruct $\chi_F(u)$ is illustrated in Fig.~\ref{fig:circuit}\textbf{a}, and their implementation based on our NMR device is shown in Fig.~\ref{fig:circuit}\textbf{b} and 2\textbf{c} for the $F$ and $B$ process, respectively.
 The completion of the protocol, which requires the exploitation of the natural coupling $\hat{\cal H}_J=2\pi J\hat\sigma^H_z\hat\sigma^C_z$ (with $J$ the coupling rate) between the $^1$H and $^{13}$C nuclear spins (cf. Fig.~\ref{fig:circuit}), encodes the characteristic function in Eq.~\eqref{eq:workcharfunc}  in the coherences of the final $^{1}$H state as $\operatorname{Re}[\chi(u)]=2 \langle \hat\sigma^H_x \rangle$ and $\operatorname{Im}[\chi(u)]=2 \langle \hat\sigma^H_y \rangle$ (cf. Appendix). This shows that the full form of $\chi(u)$ can be obtained from the $x$ and $y$ component of the $^1$H transverse magnetisation, a quantity that is straightforwardly accessed in our NMR setup.

The experiments were performed for states with different initial temperatures, sampling the characteristics function at $17.9\text{ kHz}$ rate. The interaction time $s$ in {Fig.~\ref{fig:circuit}{{\bf a}-{\bf c}} was varied through $360$ equally-spaced values for both the $F$ and $B$ process. Each realisation corresponds to an independent experiment with an average over an ensemble of molecules. The time-evolution of the measured transverse magnetisation, for the $F$ process and two different values of $T$, is shown in Fig.~\ref{fig:mag}. The amplitude of the oscillations of $\text{Re}[\chi_{F}(u)]$ (proportional to the the $x$-component of the magnetisation) is approximately the same for all temperatures (apart from a small decay in time), while a clear decrease can be seen in the imaginary part (the $y$-component of the magnetisation) as the temperature increases. In the limit of high temperatures (the maximum-entropy state), this imaginary part reduces to almost zero, as expected theoretically. 


The sample is processed in an environment at room temperature. However, the experimental data acquisition time (for each initial thermal state), which vary from $0.1$ms to $327$ms, is much smaller than the thermal relaxation time, which in NMR is associated with the spin-lattice relaxation occurring in a characteristic time $T_{1}$. In our experiment, we have measured $(T_{1}^H, T^C_1) \approx (7.36,10.55)$s. Transverse relaxation at the characteristic times $(T^H_{2},T^C_{2})\approx(4.76,0.33)$s affects, in principle, the coherences of both the system and the ancilla state. Nevertheless, the characteristic dephasing time on $^1$H is longer than the acquisition time. The situation for the $^{13}$C spin is somewhat more complicated due to the much shorter value of $T^C_2$. However, the diagonal nature of the initial equilibrium state of $^{13}$C, together with the fact that the system-ancilla coupling commutes with the map responsible for the dephasing of its nuclear spin state, leads us to claim that, as far as we only perform measurements on the ancilla, the acquired data is unaffected by the system's transverse relaxation. We observe an exponential decay during the time evolution of the magnetisation in Fig.~\ref{fig:mag}, which is mainly due to transverse relaxation. This decay only limits how long we can track the characteristic function osculations, which will ultimate bound the precision in the Fourier spectrum of the characteristic function. 

The work distribution of the experimental process is obtained from the inverse Fourier transform of $\chi(u)$. For each value of $T$ we observe well-defined peaks in the corresponding $P^\alpha(W)$ (cf. Fig.~\ref{fig:mag}{\bf e}-{\bf f}), associated to one of the four possible transitions illustrated in the lower panels of Fig.~\ref{fig:process}. {The amplitudes of the two peaks located in the $W/h<0$ semi-axis in Fig.~\ref{fig:mag}{\bf e}-{\bf f} are proportional to the excited-state population in the initial thermal state, which increases with temperature. The peaks placed in the $W/h>0$ semi-axis are instead proportional to the ground-state population, which decreases with $T$. This confirms that the conditional probabilities are independent of temperature, as it should be. 
Indeed, the $p^\tau_{m|n}$ are fixed exclusively by the quench. From the experimental data, we have estimated $p^\tau_{1|1}\approx 0.71 \pm 0.01$, $p^\tau_{0|0}\approx 0.69\pm0.01$, and $p^\tau_{0|1}\approx p^\tau_{1|0}\approx 0.71 \pm 0.01$ for both the $F$ and $B$ process. Interestingly, this provides evidence of the validity of the micro-reversibility hypothesis~\cite{campisi}. Moreover, the identities $p_{0\mid0}^{\tau}=p_{1\mid1}^{\tau}$ and $p_{1\mid0}^{\tau}=p_{0\mid1}^{\tau}$, which are valid for unital processes, can be experimentally verified from the independence of ${\rm Re}[\chi(u)]$ of $T$ (see Appendix for details).} 

\begin{figure}[t]
\includegraphics[width=\columnwidth]{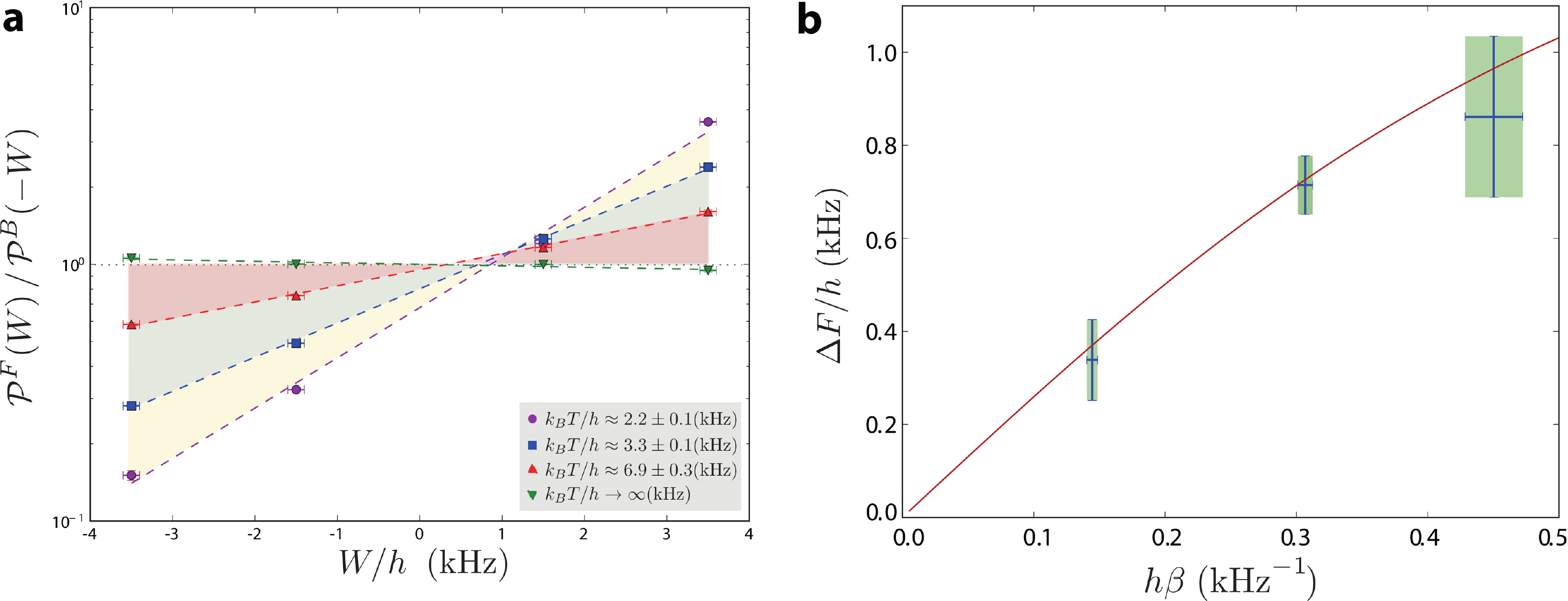}
\caption{{Verification of Tasaki-Crooks relation at the full quantum regime}. Panel {\bf a}: The ratio $P^F(W)/P^B(-W)$ is plotted in logarithm scale for the four peak frequencies in the work distribution. Panel {\bf b}: Mean values and uncertainties for the free energy variation $\Delta F$ and the inverse of the temperature $\beta$ obtained by a linear fit of the data corresponding to $T>0$ used in Panel {\bf a} to verify the Tasaki-Crooks relation. The full red line represents the theoretical expectation, $\Delta F=\frac{1}{\beta}\ln{\left(\frac{\cosh{(\beta \nu_1)}}{\cosh{(\beta \nu_2)}}\right)}$ achieved by calculating explicitly the partition functions.}
\label{fig:crooks}
\end{figure}

The experimentally reconstructed work distributions for forward and backward processes can now be used to verify important fluctuation relations such as the Jarzynski equality or the Tasaki-Crooks relation. It should not be overlooked that the processes that have been implemented in our experiments are indeed genuinely quantum mechanical, being embodied by Hamiltonians consisting of non-commuting terms. As such, the programme of experimental verifications that we report here is an important step towards the assessment of out-of-equilibrium dynamics in quantum systems subjected to a time dependent  process. 

We start computing the ratio $P^F(W)/P^B(-W)$ and use it to verify the Tasaki-Crooks relation
\begin{equation}
\label{TS}
\ln\left(\frac{P^F(W)}{P^B(-W)}\right)=\beta(W-\Delta F),
\end{equation}
where $\Delta F$ is the net change in free energy of the system resulting from the process, and which can be determined as~\cite{Silva,Dorner2} $\beta\Delta F=-\ln(Z_\tau/Z_0)$. We plot the left-hand side of Eq.~\eqref{TS} using a linear-logarithmic scale in Fig.~\ref{fig:crooks}\textbf{a}, for the values of $T$ that we have probed experimentally. The trend followed by the data-sets associated with each temperature is in very good agreement with the expected linear relation, thus confirming the predictions of the Crooks theorem. As noticed in Ref.~\cite{Dorner}, this approach can also be employed to build a high precision out-of-equilibrium thermometer which is able to capture tiny temperature variations ($\simeq 5 \pm 5\%$~nK in our experiment). The horizontal error bars in the points shown in Fig.~\ref{fig:crooks}\textbf{\bf a} are associated with the Fourier spectral linewidth, which depends on the number of oscillations resolved during the total data-acquisition time. The propagated vertical  error bars are smaller than the symbols size (see Appendix for details about the error analysis). The point at which $P^F\left(W\right) = P^B\left(-W\right)$ can be used to determine the value of $\Delta F$ experimentally. In Fig.~\ref{fig:crooks}\textbf{b}, we show $\beta$ and $\Delta F$ obtained from a linear fitting according to the aforementioned strategy. The relative error in determining $\Delta F$ arises from the fact that the actually measured parameter is the product $\beta\Delta F$.

We are now in an ideal position for the verification of the Jarzynski identity at a full quantum regime, which we here perform using three different approaches. First, we use the formulation of the Jarzynski equality~\cite{Jarzynski} $\langle e^{-\beta W}\rangle=e^{-\beta\Delta F}$ where, to fix the ideas, the average is taken over $P_F(W)$ and is determined through the relation $\langle e^{-\beta W} \rangle = \chi(u=i\beta)$, obtained by analytical continuation of the characteristic function, and making use the experimental data on the transverse magnetization of the ancillary $^{1}$H nuclear spin. Second, we use the linear fit of the Tasaki-Crooks relation, thus combining forward and backward processes. Finally, we have 
used the equivalent Jarzynski relation~\cite{Silva,Dorner2} $Z_\tau/Z_0=e^{-\beta\Delta F}$, calculating theoretically the ratio between the partition functions before and after the quench. Table~I shows the mutual agreement among these approaches: quite clearly they provide mutually consistent results within the respective associated uncertainties (the uncertainties in the theoretical results used in our third approach arise from the uncertainty in temperature of the thermal state), thus providing a statistically significant verification of the Jarzynski equality. It is remarkable that the left and right-hand side of the equivalent equalities used for the assessment of the Jarzynski relation have been determined experimentally in completely independent ways.

\begin{table}[t]
\includegraphics[width=\columnwidth]{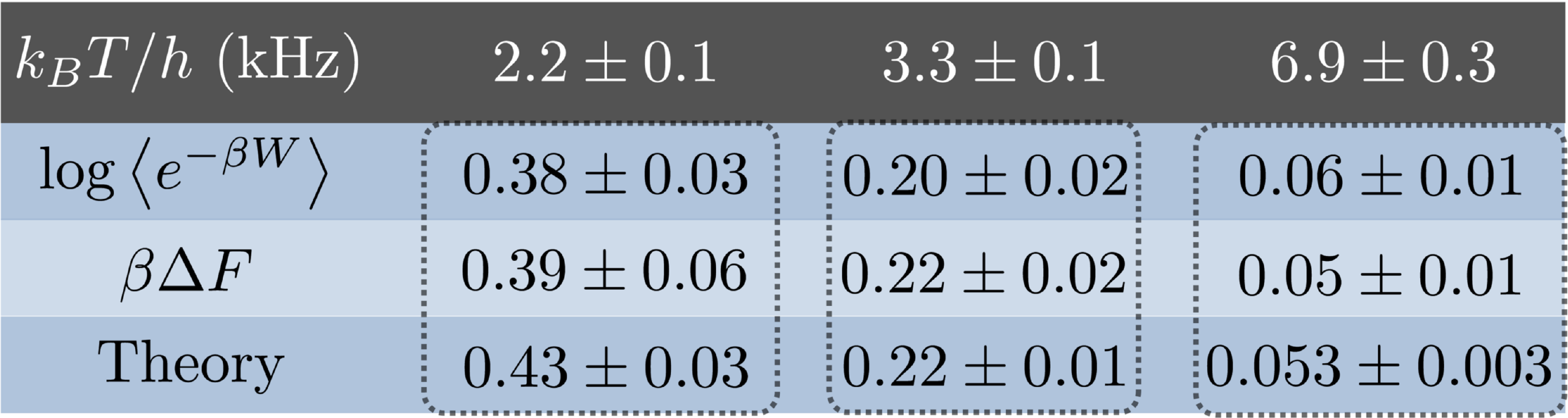}
\caption{{Verification of Jarzynski identity at the full quantum regime}. We report the experimental values of the left and right-hand side of the Jarzynski identity, measured in two independent ways for three choices of temperature, together with the respective uncertainties (see main text for details). The experimental results are compared to the theoretical prediction coming from the ratio $\ln(Z_\tau/Z_0)$. The agreement among the three values is consistent for the three values of temperature probed in our experiment.}
\label{fig:jaarzynski}
\end{table}

We have explored experimentally the statistics of work following  a quantum non-adiabatic process in a spin-$1/2$ system using an ancilla-based interferometric approach adapted to NMR technology. Our original way to reconstruct the work characteristic function has allowed us to address, for the first time to our knowledge, fluctuation relations at the full quantum level and thus go significantly beyond the current experimental state-of-the-art, which was so far constrained to the classical domain. Our results demonstrate the feasibility of an experimental study on the thermodynamics of out-of-equilibrium quantum systems and paves the way to its extension to the quantum many-body context~\cite{Dorner2,Dudu}.

\acknowledgments
We thank M. Campisi, L. Celeri, S. F. Huelga, K. Modi, G. M. Palma, M. B. Plenio, F. L. Semiao, D. O. Soares-Pinto, and V. Vedral for valuable discussions. We acknowledge financial support from CNPq, CAPES, FAPERJ and FAPESP. This work was performed as part of the Brazilian National Institute of Science and Technology for Quantum Information (INCT-IQ). LM is supported by the EU through the Marie Curie Action. M.P. acknowledges the UK EPSRC for a Career Acceleration Fellowship and a grant awarded under the ``New Directions for Research Leaders'' initiative (EP/G004579/1), the CNPq ``Ci\^{e}ncia sen Fronteiras'' programme for a grant under the ``Pesquisador Visitante Especial'' initiative (grant nr. 401265/2012-9), and the Alexander von Humboldt Stiftung. 
L.M. and M.P. are grateful to the Universidade Federal do ABC and the CBPF labs for the kind hospitality during the early stages of this work.

\section*{\large APPENDIX}

\subsection{Description of the experimental setup}

The sample employed in our experiments is made of 50 mg of 99\% $^{13}$C-labeled CHCl$_3$ diluted in 0.7 ml of 99.9\% deutered Acetone-d6, in a flame sealed Wildmad LabGlass 5 mm tube.  The experiments were performed at $T'=25^{\circ}$Celsius using a Varian $500$~MHz Spectrometer and a double-resonance probe-head equipped with a magnetic field gradient coil. The nuclear spins were manipulated through suitably designed sequences of rf-fields. The density operator of the $^1$H-$^{13}$C spin pair can be described, in the high temperature expansion~\cite{Ivan, Jones}, as  $\rho \approx{\openone}/{4}+\varepsilon\Delta\rho$, where $\Delta\rho$ is the deviation matrix and $\varepsilon=\frac{\hbar\gamma B_0}{4k_{B}T'} \approx 10^{-5}$ is the ratio between magnetic and thermal energies with $\gamma$ the gyromagnetic ratio and $B_0 \approx 11.75$~Tesla the intensity of a strong static magnetic field (whose direction is taken to be along the positive $z$ one).  The $^{1}$H and $^{13}$C nuclear spins precess around $B_0$ with Larmor frequencies ${\omega_H}/{2\pi} \approx 500$~MHz and ${\omega_C}/{2\pi} \approx 125$~MHz, respectively. The deviation matrix $\Delta\rho$ is the accessible part of the system state in NMR experiments and all rf-field manipulations of the system state act only on it. This system encompasses all features of quantum dynamics, such as coherent evolutions and interference~\cite{Ivan, Ladd, Auccaise1}. It also supports general quantum correlations~\cite{Soares-Pinto, Auccaise2}. As discussed in the main text and in one of the following Subsections, in the present experiment the $^{1}$H-$^{13}$C spin pairs were initially prepared in a diagonal state of the so-called computational basis, $\{\ket{0_H0_C},\ket{0_H1_C},\ket{1_H0_C},\ket{1_H1_C}\}$, with the $^{13}$C spin populations weighted by the thermal ensemble. This procedure maps the deviation matrix ($\Delta\rho$) into a pseudo-equilibrium state at a given temperature $T$, as far as the  $^{13}$C spins are involved. In this way we can simulate thermal state at arbitrary temperatures.}

\subsection{Characterization of the initial equilibrium state}

\begin{figure}[b]
\includegraphics[width=0.75\columnwidth]{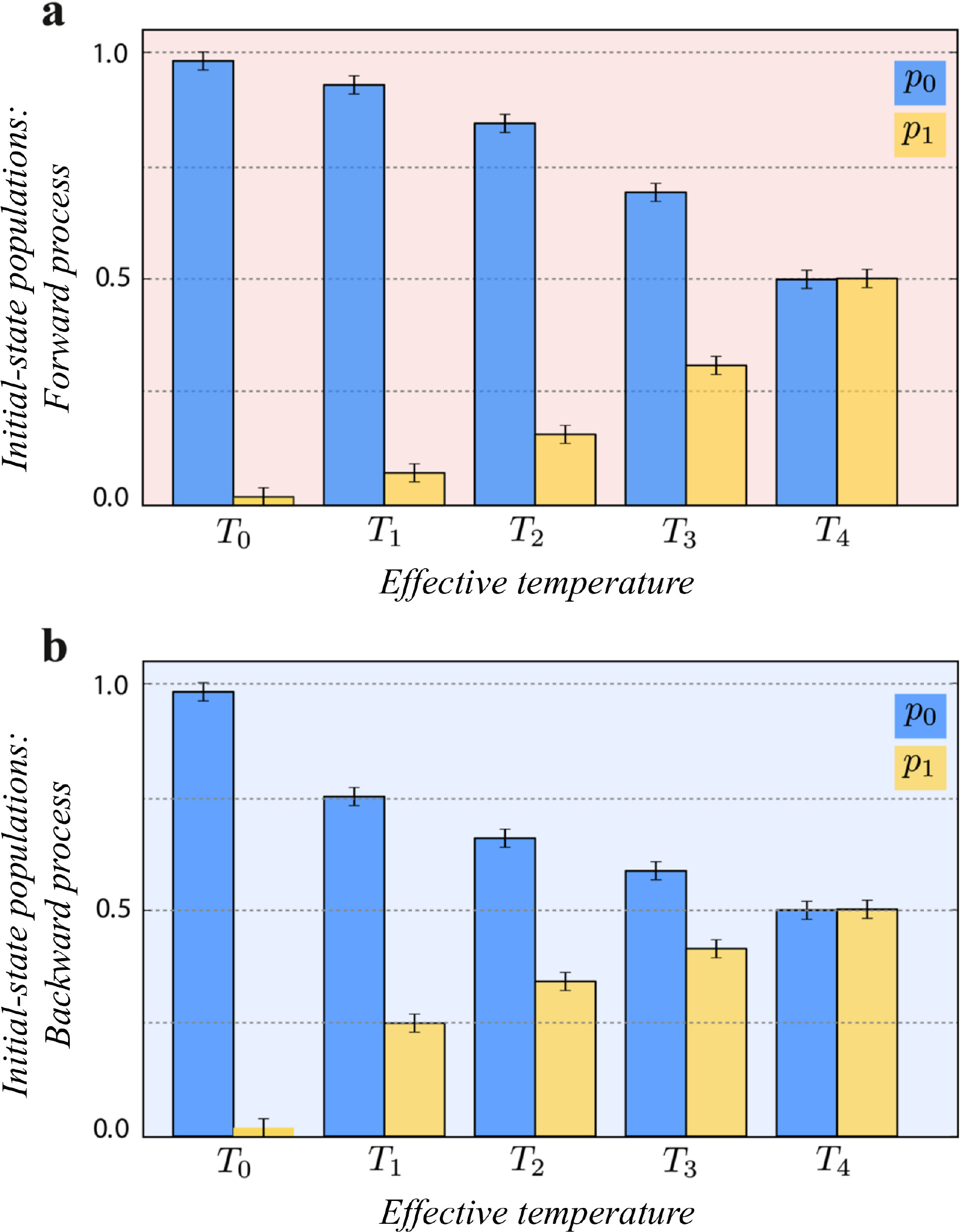}
\caption{{Populations of the initial state of $^{13}$C}. Populations of the ground-state (blue columns) and 
excited-state (orange columns) of the initial states, prepared for the forward (panel \textbf{a}) and backward (panel \textbf{b}) processes at temperature $T_k$ (see Table I for the explicit value of the initial state temperature).}
\label{fig:pops}
\end{figure}

\begin{table}[t]
\begin{tabular}{cccccc}
\hline 
 & \multicolumn{2}{c}{Excited-state population ($p_{1}^{0}$)} & \multicolumn{3}{c}{$k_BT / \hbar$ (kHz)} \\
 & Forward & Backward & Forward & Backward & Average \\
\hline 
\hline 
$T_{0}$ & $0.02\pm0.02$ & $0.02\pm0.02$ & $\to0$ & $\to0$ & $\to0$\\
$T_{1}$ & $0.07\pm0.02$ & $0.25\pm0.02$ & $2.0\pm0.1$ & $1.8\pm0.1$ & $1.9\pm0.1$\\
$T_{2}$ & $0.16\pm0.02$ & $0.34\pm0.02$ & $3.0\pm0.1$ & $3.1\pm0.2$ & $3.1\pm0.2$\\
$T_{3}$ & $0.31\pm0.02$ & $0.41\pm0.02$ & $6.2\pm0.4$ & $5.8\pm0.7$ & $6.0\pm0.7$\\
$T_{4}$ & $0.50\pm0.02$ & $0.50\pm0.02$ & $\to\infty$ & $\to\infty$ & $\to\infty$\\
\hline 
\end{tabular}
\caption{{Characterization of the initial states of the $^{13}$C nuclear spins obtained from quantum
state tomography}. The state is very close to a pseudo-thermal state (for
the $^{13}$C nuclear spin); the inverse temperature of the actually prepared state is estimated 
from the excited state population as $T=2\nu_{k}\log\left(\frac{p_{1}^{0}}{1-p_{1}^{0}}\right)$,
where $k=1,2$ for the forward and backward processes, respectively.  The temperatures for the forward and backward processes are equivalent within the experimental errors.}
\end{table}

In our experiments, the $^{13}$C nuclear spins of our sample are initially prepared in a pseudo thermal state at temperature $T$, 
whereas the $^{1}$H ones are initialised in the ground state. 
We have characterised such states by means of full quantum state tomography~\cite{Ivans}. The population distribution for $^{13}$C is shown in Fig.~{\bf \ref{fig:pops}} for both the forward ($F$) and backward ($B$) process. From such information, we have been able to estimate the effective temperature of the initial pseudo thermal state of the $^{13}$C nuclear spins, which are reported in Table~1. The initial Hamiltonian of the forward process [$\hat{\cal H}^F (0)$] has a large energy gap between the fundamental and excited state of the $^{13}$C nuclear spins, which implies that the excited-state population increases more slowly with temperature compared to the backward process, whose initial Hamiltonian has a smaller energy gap (cf. Fig. 1 in the main text). The unequal occupation probabilities in the initial state of the \textbf{F} and \textbf{B} process are due to such different energy gaps.

\subsection{Characterization of the process}

The $F$ and $B$ process that we have implemented experimentally have been described in the main text. Here we provide a more in-depth 
characterisation of the transformations resulting from the quenches that we have considered. 

\begin{figure}
\includegraphics[width=0.95\columnwidth]{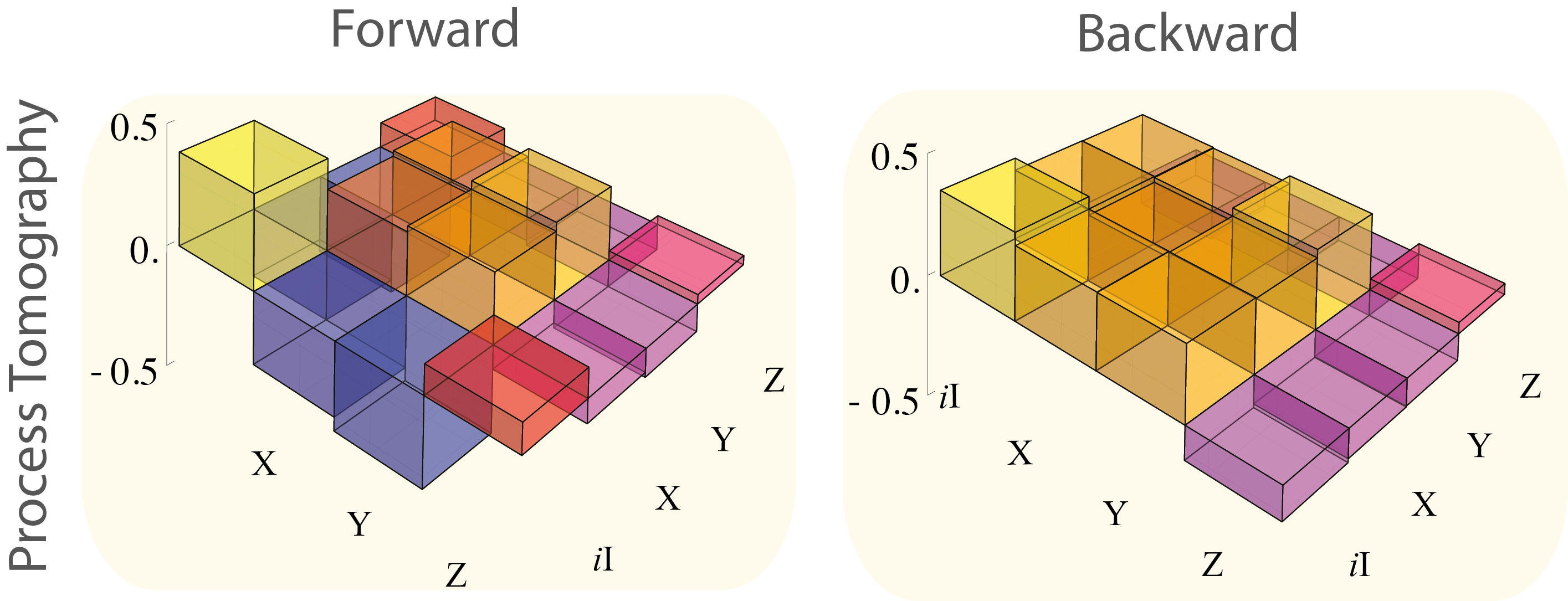}
\caption{{Forward and backward process}. The real part of the entries $\xi_{k,l}$ of the experimental
quantum process tomography matrix (the imaginary part is close to zero in the chosen operator basis). 
The left (right) panel is relative to the forward (backward) process.}
\label{fig:process1}
\end{figure}

Needless to say, experimental imperfections and non-idealities prevent the realisation of a 
perfectly unitary evolution. In our implementation of quenched dynamics, there are two main error sources. 
The first one is the inhomogeneity in the intensity of the rf-pulse across the sample. This causes a deviation
from unitary evolution. 
The second source is embodied by the $^1$H-$^{13}$C coupling, which might establish initial correlations between 
the nuclear spins of the two atomic species. In order to determine how close the processes that we have considered are to ideal unitary evolutions, we employ the diagnostic instrument provided by quantum process tomography (QPT)~\cite{Nielsens}. A general evolution undergone by the initial state of the $^{13}$C spin is provided by the quantum map 
\begin{equation}
\mathcal{E}(\rho^{C})=\sum_{k,l=0}^{3} \xi_{kl}\hat\sigma_{k}^{C}\rho^{C}\hat\sigma_{l}^{C\dagger},
\end{equation}
with $\hat{\bm\sigma}^C=(\hat\sigma^C_0,\hat\sigma^C_x,\hat\sigma^C_y,\hat\sigma^C_z)$, $\hat\sigma^C_{0}
\equiv i\hat\openone^C$, and ${\bm \xi}$ the matrix of the process, which completely identifies the map. The basis ($\hat{\bm\sigma}^C$) differs from the one picked in Ref.~\cite{Nielsens}, our choice being motivated by the fact that any unitary process corresponds to ${\bm\xi}\in {\cal M}(4,{\mathbb R})$. To implement QPT experimentally, we have prepared the set of input pseudo-pure states 
$\{|0\rangle,|1\rangle,(|0\rangle +| 1\rangle)/\sqrt{2},(|0\rangle+i|1\rangle)/\sqrt{2}\}$~\cite{Nielsens}, subjected each of them to the process to characterise, and finally reconstructed the output density matrix using quantum state tomography~\cite{Ivans, Joness}. Numerical post processing of the acquired data enables the final estimate of the entries of ${\bm\xi}$, which are shown (for the $F$ and $B$ processes) in Fig.~\ref{fig:process1}. 

Unital processes, i.e. those such that $\mathcal{E}(\openone) = \openone$, embody a very important class of maps encompassing unitary ones. For both the forward and backward dynamics implemented in our experiment, we estimated the best general 
unitary and unital process that describe the system's evolution. As a measure 
of the distance between two arbitrary processes ${\cal E}_{1,2}$, we have used the trace distance~\cite{Nielsens} \hbox{$\delta(\rho^{C},{\cal E}_1,{\cal E}_2)=\frac{1}{2} \mathrm{Tr}\hspace{1pt}|{\cal E}_1(\rho^{C})-\mathcal{E}_2(\rho^{C})|_1$} with $|{\bm x}|_1={\rm Tr}\sqrt{{\bm x}^\dag{\bm x}}$ the norm-1 of a matrix ${\bm x}$. Moreover, in order to wash out any dependence on the specific initial state and account for a worst-case scenario, we have considered $\delta({\cal E}_1,{\cal E}_1)=\max_{\rho^C}\delta(\rho^{C},{\cal E}_1,{\cal E}_2)$ over all choices of initial state. 
Such worst case trace distance between the ideal process ${\cal E}_1=\hat{\cal U}\rho^{C}\hat{\cal U}^{\dagger}$ and the experimentally reconstructed one ${\cal E}_2={\cal E}$ is found to be $\simeq3.3\%$ ($7.7\%$) for the $F$ ($B$) process. The experimental map is also almost perfectly unital, as the trace distance for the identity evolution, $\delta(\openone,{\cal E}_1,{\cal E}_2)\simeq0.9\%$ ($3.9\%$) 
for the forward (backward) process. 
Therefore, notwithstanding the ubiquitous influence of experimental imperfections and environmental mechanisms discussed in the main text, our setup operates very close to an ideal unitary evolution. 

\subsection{Experimental procedure to reconstruct the work distribution}

After the initial state preparation, a quench of the magnetic field drives the $^{13}$C nuclear spins out-of-equilibrium. 
As described in the main text, the statistics of the corresponding work is reconstructed from a series of independent 
measurements of the transverse magnetization of the $^1$H nuclear spins, folowing the interferometric strategy proposed in Refs.~\cite{Dorner,Mazzola}.

The conditional operations $\hat G_1$ and $\hat G_2$ (Eq. (3) of the main text) needed for this task are equivalent (up to a global phase) 
to an appropriate sequence of two-spin operations generated by the coupling Hamiltonian $\hat{\cal H}_J=2\pi J \hat\sigma^H_z \hat\sigma^C_z$, local 
rotations on the $^{13}$C nuclear spins and a spin-flip of the $^1$H one, as sketched in Figs. 2\textbf{b} and 2\textbf{c} of the main text. 

Both the initial state of the sample and the $\hat{\cal H}_J$ coupling are diagonal in the computational basis, whereas the Hamiltonian responsible for the quench 
on the system ($\hat{\cal H}^{F}(t)$ and $\hat{\cal H}^{B}(t)$ defined in the main text) are {\it transversal},  both for the $F$ and the $B$ processes. Before the quench, 
we suitably rotate the $^{13}$C nuclear spin to the eigenbasis of $\hat{\cal H}^{\alpha}(0)$ ($\alpha=F,B$), as described in Figs. 2\textbf{b} and 2\textbf{c} of the main text. 
This corresponds to an abridged implementation of $\hat G_1$. Likewise, after the quench, we rotate the system from the eigenbasis of $\hat{\cal H}^{\alpha}(\tau)$ back to the computational basis for the application of the second $\hat{\cal H}_J$ coupling, which is equivalent to the controlled operation $\hat G_2$. 
Such rotations are identified by $L$ and $K$ in Figs. 2\textbf{b}, \textbf{c} of the main text. 
The two $\hat{\cal H}_J$  couplings act for different time-lengths so to compensate the difference in energy gap between initial
and final Hamiltonian. The first free evolution is 
2.5-fold longer (shorter) than the second one for the $F$ ($B$) process.

A detailed calculation of the operation performed by the quantum circuit for the reconstruction of the characteristic function leads to the following expression for the magnetization of $^1$H
\begin{align}
 M^{\alpha}\left(s\right) &= \frac{1}{2}\left[p_1^{\tau,\alpha} p_{0\mid 1}^{\tau,\alpha} e^{-i\left(\nu_1+\nu_2\right)2\pi s}  
 + p_1^{\tau,\alpha} p_{1\mid 1}^{\tau,\alpha} e^{\mp i\left(\nu_1-\nu_2\right)2\pi s}\nonumber \right.\\ &
 + p_0^{\tau,\alpha} p_{0\mid 0}^{\tau,\alpha} e^{\pm i\left(\nu_1-\nu_2\right)2\pi s}  
 \left.+ p_0^{\tau,\alpha} p_{1\mid 0}^{\tau,\alpha} e^{i\left(\nu_1+\nu_2\right)2\pi s}\right].
\label{eq:mag}
 \end{align}

\begin{figure}
\includegraphics[width=0.95\columnwidth]{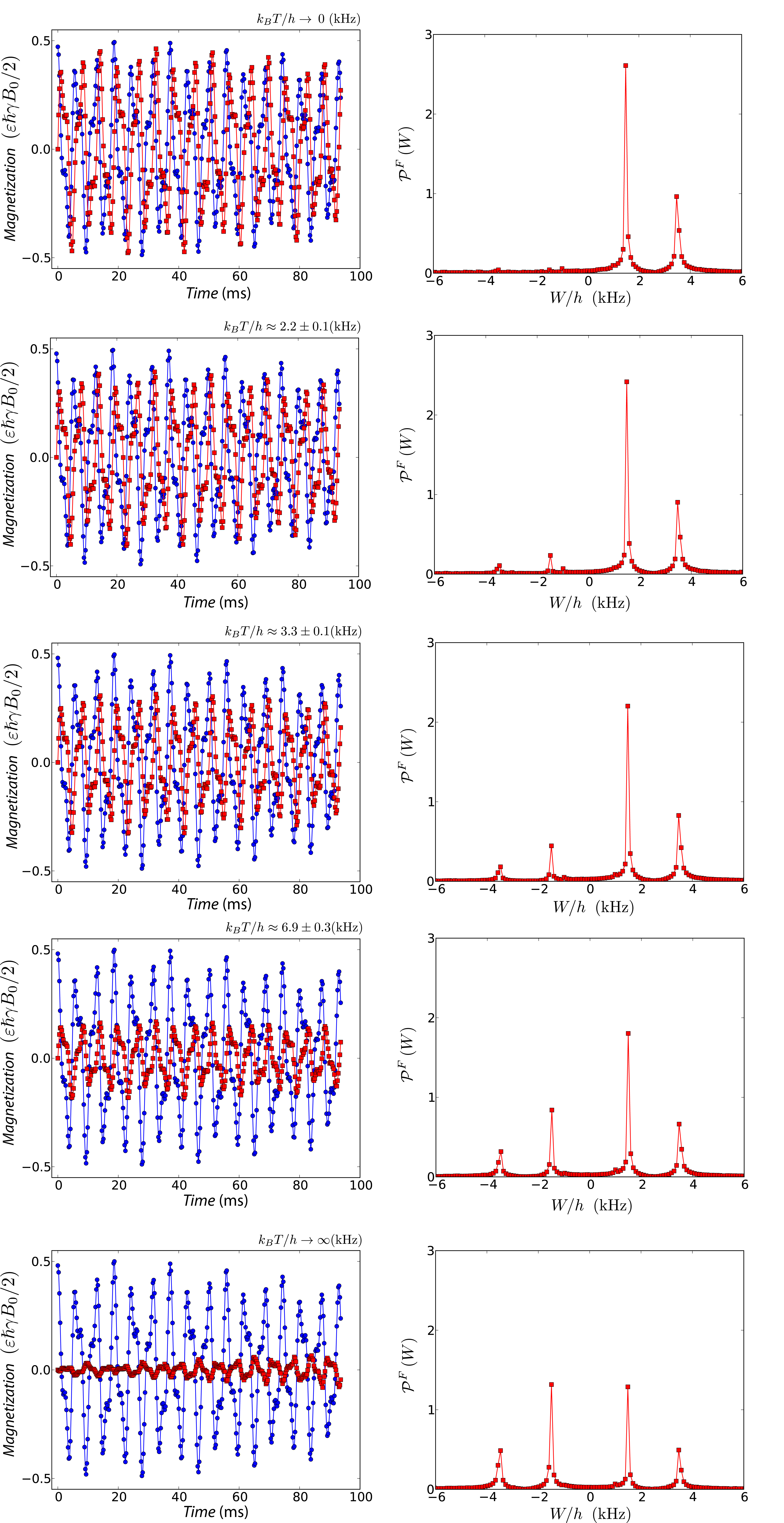}
\caption{Transverse magnetisation (left column) and work probability distribution for the $F$ process and five different values of the initial temperature of the $^{13}$C nuclear spins state.}
\label{fig:bigtableFor}
\end{figure}
\begin{figure}
\includegraphics[width=0.95\columnwidth]{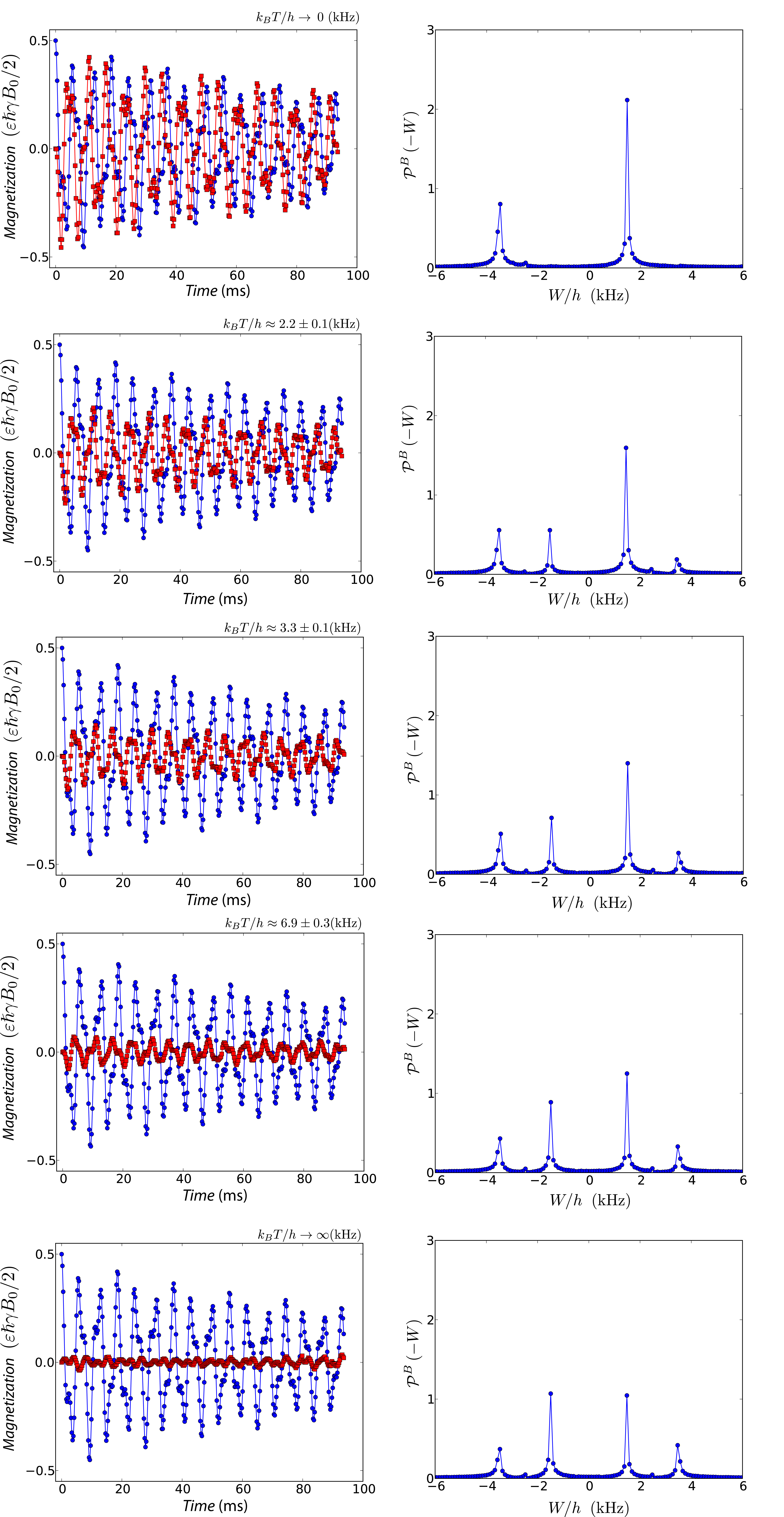}
\caption{Transverse magnetisation (left column) and work probability distribution for the $B$ process and five different values of the initial temperature of the $^{13}$C nuclear spins state.}
\label{fig:bigtableBack}
\end{figure}

Experimentally, we have applied the protocol for the reconstruction of the characteristic function to the case of initial pseudo-thermal states of the $^{13}$C nuclear spins corresponding to each of the temperatures reported in Table~II, for both processes. In Fig.~\ref{fig:bigtableFor} and ~\ref{fig:bigtableBack} we report the experimentally acquired data for the corresponding transverse magnetisations and their inverse Fourier transforms. In the main text we reported and discussed only the cases corresponding to two of such temperatures. 

\subsection{Micro-reversibility}

Having characterised fully the physical process implemented in our experiment, we can reformulate the conditional probabilities $p^{\tau,\alpha}_{m|n}$  in terms of the entries of the process matrix ${\bm\xi}$ as (we have explicitly introduced the label $\alpha=F,B$ to distinguish the values associated with the forward and backward process)
\begin{equation}
p_{m\mid n}^{\tau,\alpha} = \sum_{k,l=0}^{3} \xi_{kl}^{\alpha} \langle m^\alpha(\tau)| \hat{\sigma}_k^{C} \mid n^{\alpha}(0) \rangle
\langle n^{\alpha}(0) \mid \hat{\sigma}_l^{C\dagger} \mid m^{\alpha}(\tau) \rangle
\end{equation}
with $\ket{n^{\alpha}(0)}$ and $\ket{m^{\alpha}(\tau)}$ eigenstates of $\mathcal{H}^{\alpha}(0)$ and $\mathcal{H}^{\alpha}(\tau)$.
Comparing $p^{\tau,F}_{m|n}$ and $p^{\tau,B}_{m|n}$, using the corresponding expressions valid  for the ideal process and the expression for the backward time propagator given in the main letter, we obtain $p_{n\mid m}^{\tau,B} = p_{m\mid n}^{\tau,F}$, which states the micro-reversibility of the process. 
It is well-known that unitality is the quantum counterpart of classical microreversibility~\cite{Zanardi}. It is thus not surprising that, having proven in Sec. C the almost full unital nature of the processes at hand, microreversibility is also retrieved and verified here. 
We can actually verify microreversibility from the dynamics of the real part of the magnetization, which according to Eq.~(\ref{eq:mag}) is
\begin{align}
{\rm Re}\left(M^{\alpha}\left(s\right)\right) & =\frac{1}{2}\left(p_{1}^{\tau,\alpha}p_{0\mid1}^{\tau,\alpha}+p_{0}^{\tau,\alpha}p_{1\mid0}^{\tau,\alpha}\right)
\cos[2\pi(\nu_{1}+\nu_{2})s] \nonumber \\
 & +\frac{1}{2}\left(p_{1}^{\tau,\alpha}p_{1\mid1}^{\tau,\alpha}+p_{0}^{\tau,\alpha}p_{0\mid0}^{\tau,\alpha}\right)\cos[2\pi (\nu_{1}-\nu_{2})s].
\end{align}
The experimental observation that the real part of the magnetization is independent of the temperature of the initial thermal state (see Fig. 2 of the main text, where such feature is well evident) implies that $p_{0\mid 0}^{\alpha} = p_{1\mid 1}^{\alpha}$ and $p_{0\mid 1}^{\alpha} = p_{1\mid 0}^{\alpha}$. Therefore this dynamical behaviour also verifies micro-reversibility.

\subsection{Error Analysis}


The magnetization curves measured in our experiment (and shown in Fig. 3 of the main text for both $F$ and $B$ process), 
are proportional to the Fourier transform of work probability distribution and, as seen from Eq.~(\ref{eq:mag}), can be cast into the form
\begin{equation}
M_{theory}(t)=e^{-\gamma_d t}\sum_{k=1}^{4}\alpha_{k}e^{-i\omega_{k}t}\ ,\label{eq:form}
\end{equation}
where $\gamma_d$ is a decay rate, caused mainly by field inhomogeneity across 
the sample, and $\omega_{k}$ are the frequencies of the four peaks appearing in
Fig. 3{\bf e} and 3{\bf f} of the main text. Given our experimental parameters, from the leftmost to to the rightmost peak, the expected values of such frequencies are $\approx\left( -3.5,-1.5,+1.5,+3.5\right)$kHz. The exponential decay term has been added to the magnetisation in Eq.~\eqref{eq:mag} so as to account for experimental imperfections.
We can now estimate the best values of the coefficient entering Eq.~\eqref{eq:form} and their associated uncertainties using standard statistical analysis tools, and get the best fit of the experimental data.

Clearly, the magnetisation curve observed in our experiment depends linearly on eight real parameters (the four complex amplitudes
$\alpha_k$) and non-linearly on three others (the frequencies $\omega_k$ and the decay rate $\gamma_d$).
The decay rate is not of our direct interest, given the phenomenological nature of such parameter. On the other hand, 
the frequencies $\omega_k$ are related to the gaps in the initial and final Hamiltonians and the $\alpha_k$'s 
are linked to the transition probabilities, and are the ones we are most interested in.


The least-squares method has been applied to 10 experimentally reconstructed magnetization curves (we have sampled $5$ different temperatures for both $F$ and $B$ process, as reported in Table~II, Fig.~\ref{fig:bigtableFor} and Fig.~\ref{fig:bigtableBack}). The resulting optimal estimate of $\omega_k$'s are consistent across all the data sets and in excellent agreement with the theoretical expectations based on the values of $\nu_{1,2}$ stated in the main text. The variances of all fitting parameters are much smaller than the perceived line-widths in Fig. 3{\bf e} and {\bf f}, and are caused only by a limited resolution of the Fourier transform.  We also checked that the decay rate $\gamma_d$ in the fitting function does not depend on the temperature of the initial state, although it depends on the process, being $4$ times larger in the $B$ process than in the $F$ one (it corresponds to 20\% of attenuation of magnetization data for the $B$ dynamics, as can be seen in Fig. 3 of the main text).


\subsection{Verification of the Tasaki-Crooks relation}

In Fig. 4\textbf{a} of the main text, the ratio of the probability distributions for the $F$ and $B$ process are plotted in logarithmic scale.
Although the evaluation of the uncertainties associated with such data requires many steps of error propagation, the vertical error bars are all smaller than the 
size of the data point in the Figure. The least squares method has been used to fit these points to a linear relation, and we have estimated both the angular coefficient (which is found consistent with the inverse temperature $\beta$), and the vertical shift from the origin, which provides an estimate of $\beta\Delta F$. The temperatures estimated from the Tasaki-Crooks relation were used to label the lines in Fig. 4\textbf{a}. The theoretical expectation for the relationship between $\beta$ and $\Delta F$ 
\begin{equation}
 \Delta F = \frac{1}{\beta} \log\left( \frac{\cosh\left(\beta\nu_1\right)}{\cosh\left(\beta\nu_2\right)} \right) .
 \label{eq:theory}
\end{equation}
was finally plotted in Fig. 4\textbf{b}, showing remarkable agreement with the experimental data. 

\subsection{Verification of the Jarzinski equality}

Dividing the magnetization fitting function Eq.~(\ref{eq:form}) by $\sum_{k=1}^{4}\alpha_{k}$, we obtain the work characteristic function $\chi\left(u\right)$. For the verification of 
the Jarzynski identity, we are interested in the value of this function in $i\beta$. This particular value of the characteristic function is estimated by analytical continuation of the experimental data 
[cf. Eq. (\ref{eq:form})]. However, the inverse temperature $\beta$ (obtained from the Tasaki-Crooks relation)
itself carries an uncertainty due to the indeterminacy associated with the temperature of the initial states of our sample. We have thus used a Monte Carlo method, sampling values of  the parameters set $\left(\gamma_d,\{\omega_{k}\},\left\{ \alpha_{k}\right\} \right)$ with a multivariate Gaussian distribution having widths determined by the variances corresponding to such parameters. We have also sampled $\beta$ with an independent Gaussian distribution (width determined by the corresponding experimental uncertainty). We have thus estimated mean and standard deviation of the distribution of values of $\chi\left(i\beta\right)$ that we have gathered. 

The expected theoretical value of $\beta\Delta F$ given by Eq.~\eqref{eq:theory} (including the uncertainty due to the initial temperature of the sample) has finally been used as a milestone for our verification of the Jarzynski identity, as reported in Table~I of the main text.

\end{document}